\documentclass[aps,prl,twocolumn]{revtex4}
\usepackage{amsmath, amsthm, amssymb}
\usepackage[dvips]{graphics}
\usepackage[dvips]{graphicx}
\begin{document}
\title{On the relation between decoherence and spontaneous symmetry breaking}
\author{Jasper van Wezel$^1$, Jan Zaanen$^1$ and Jeroen van den Brink$^{1,2}$}
\affiliation{
$^1$ Institute-Lorentz for Theoretical Physics,  Universiteit  Leiden,\\
P.O. Box 9506, 2300 RA Leiden,The Netherlands\\
$^2$ Institute for Molecules and Materials, Radboud Universiteit Nijmegen,\\ 
P.O. Box 9010, 6500 GL Nijmegen, The Netherlands}

\date{\today}

\begin{abstract}
We have recently shown that there is a limit to quantum coherence in
many-particle spin qubits due to spontaneous symmetry breaking. These results
were derived for the Lieb-Mattis spin model. Here we will show that the
underlying mechanism of decoherence in systems with spontaneous symmetry
breaking is in fact more general. We present here a generic route to
finding the decoherence time associated with spontaneous symmetry breaking in
many particle qubits, and subsequently we apply this approach to two
model systems, indicating how the continuous symmetries in these models are
spontaneously broken and discussing the relation of this symmetry breaking to the
thin spectrum.  We then present in detail the calculations that lead to
the limit to quantum coherence, which is due to energy shifts in the
thin spectrum.
%
%Here we present a detailed study of this model, indicate how the continuous spin rotation symmetry is spontaneously broken in it and discuss the relation of this symmetry breaking to the thin spectrum.  We then present in detail the calculations that result into the limit to quantum coherence, which is in fact due to the thin spectrum.  
\end{abstract}

\maketitle

\section{Introduction}
Recently we have shown that spontaneous symmetry breaking imposes a
fundamental limit for the time that a large spin system can stay quantum
coherent. This universal time-scale is $t_{spon} \simeq 2\pi N \hbar / (k_B
T)$, given in terms of the number of microscopic degrees of freedom $N$,
temperature $T$, and the constants of Planck ($\hbar$) and Boltzmann
($k_B$)~\cite{Wezel05}.
We analyzed this quantum decoherence process in terms of the exactly solvable
Lieb-Mattis spin model, which is known to describe a symmetry broken
macroscopic antiferromagnet in equilibrium. Within this spin model we
investigated the dynamical reduction of quantum physics to classical behavior
via spontaneous symmetry breaking. The goal of this paper is to present a
self-contained, detailed description and explanation of the decoherence
process caused by spontaneous symmetry breaking.

The fact that spontaneous symmetry breaking can lead to decoherence on a time
scale $t_{spon}$ might come as a surprise. For a macroscopic body at room
temperature, $\hbar / (k_B T)$ is of order $10^{-14}$ seconds, which is quite
a short time. However, multiplying with Avogadro's number $N \simeq 10^{24}$,
yields $t_{spon} \simeq 10^{10}$ seconds, corresponding to a couple of
centuries. Given all other sources of decoherence for such a large macroscopic
body, this is surely not a relevant timescale. However, the mesoscopic quantum
qubits of contemporary interest are typically much smaller and the intrinsic
coherence time might be reached in the near future.  The counterintuitive
feature of this intrinsic decoherence mechanism linked to classical
equilibrium is that it starts to matter when systems become small. 

The many-particle qubits that motivate us to study decoherence due to
spontaneous symmetry breaking are realized in a number of mesoscopic solid
state systems. For instance, by engineering aluminum on a sub-micron length
scale, superconducting flux qubits and Cooper pair boxes can be
manufactured. The flux qubit is a Josephson device that can be brought into a
quantum superposition of two electrical currents: a left and a right
circulating current~\cite{Chiorescu00,Wal00}. Typically this current is
carried by $N \sim 10^6$ Cooper pairs.  A Cooper pair box on the other hand is
a superconducting island, containing $N \sim 10^8$ electrons, which can be
brought in superposition of two states with a different number of Cooper
pairs~\cite{Nakamura99,Vion02}.
Magnetic many-particle  qubits are for instance realized in molecular
nanomagnets.  Molecules with large magnetic moments can be brought into a
superposition of directions of magnetization. A well studied example is
Mn$^{12}$ acetate, a molecule that contains 12 manganese atoms, coupled
together to form a total spin of $S=10$. The molecule can be brought into a
superposition of states with $S^z =+10$ and $S^z =-10$ and coherent
Rabi-oscillations of the magnetization are observed~\cite{Friedman96}. An even
larger molecule is ferritin, that contains about 4500 Fe$^{3+}$
ions~\cite{Awschalom92}. If the total magnetic moment of a ferritin molecule
is brought into a coherent superposition, this corresponds to a superposition
of $N \sim 10^2$ spins. For these mesoscopic superconducting and magnetic
qubits the limit in coherence due to spontaneous symmetry breaking is
relevant.

This paper is organized as follows. We first introduce the general notion of
spontaneous symmetry breaking and the thin spectrum. We illustrate these
concepts by the elementary example of a harmonic crystal that breaks
translational symmetry. We then show how the presence of a thin spectrum can
cause decoherence. Again we clarify the concept by describing a theoretical
setup in which we use the crystal as a qubit. After that we switch to the more
involved example of the (non-abelian) antiferromagnet. We first recapitulate
the properties of the Lieb-Mattis model~\cite{Lieb62} and determine its
symmetric and symmetry-broken eigenstates. Subsequently we will use the model
to determine the effect of spontaneous symmetry breaking on quantum
coherence. For this purpose we devise a generic gedanken experiment by
coupling a two spin singlet state to the macroscopic Lieb-Mattis
antiferromagnet. We calculate the exact time evolution of this many-body
quantum superposition. Our main result on decoherence follows from the
evaluation of the reduced density matrix of the superposition by tracing out
the thin states. Finally we will then discuss what it reveals regarding
spontaneous symmetry breaking in more general Heisenberg-like spin
Hamiltonians.

\subsection{Spontaneous Symmetry Breaking} 
Due to the homogeneity of space, the laws of nature possess translational
invariance. This invariance implies the classical law of conservation of
momentum. Likewise, space being isotropic enforces rotational symmetry on the
laws of physics, implying conservation of angular momentum. In quantum
mechanics the power of symmetry is even greater: the translational and
rotational invariance of the laws of nature taken at face value, should imply
that any physical quantum object that obeys these laws, has translational and
rotational symmetry. However, daily experience shows that this conclusion is
nonsensical. If the universe around us and everything in it would be
translationally and rotationally invariant, it would look the same at all
places and in all directions: we would be surrounded by a "quantum mist",
while human observers should be dissolved in this quantum fog as
well. However, in the real world translational and rotational symmetry are
manifestly broken.
The fundamental difference between quantum and classical physics lies in the
role of symmetry. Dealing with an exact quantum mechanical eigenstate, all
configurations equivalent by symmetry should have exactly the same status in
principle, while in a classical state one of them can be singled out. In the
example above, given that space is translationally invariant, a quantum object
should be in an eigenstate of total momentum, being spread out with equal
probability over all of space. In the classical limit however it takes a
definite position. The explanation of this `spontaneous symmetry breaking' as
a ramification of the singular nature of the thermodynamic limit is one of the
central achievements of quantum condensed matter
physics~\cite{Anderson52}. One imagines a symmetry breaking `order parameter
field' $B$ (e.g., a potential singling out a specific position in space). Upon
sending $B$ to zero before taking the thermodynamic limit ($N \rightarrow
\infty$) one finds the exact quantum groundstate respecting the
symmetry. However, taking the opposite order of limits one finds that the
classical state becomes fact. Although the concept of spontaneous symmetry
breaking was originally introduced in the context of quantum magnetism in
solid state physics~\cite{Anderson52}, spontaneous symmetry breaking is a
general phenomenon, that is just as relevant in other fields, including
elementary particle physics and cosmology~\cite{Coleman85}.

Let us first consider how spontaneous symmetry breaking arises in a
crystalline lattice to continue in the next section with
antiferromagnets. Consider the textbook example of a harmonic crystal, with
the Hamiltonian
\begin{eqnarray}
H=\sum_{j} \frac{{\bf p}^2_j}{2 m} + \frac{\kappa}{2} \sum_{j}  \left( {\bf
  x}_j - {\bf x}_{j+1} \right)^2,
\label{eq:Xtal}
\end{eqnarray}
where $j$ labels all $N$ atoms in the lattice, which have mass $m$, momentum
${\bf p}_j$ and position ${\bf x}_j$. The harmonic potential between
neighboring atoms is parameterized by $\kappa$; it turns out that the results
on spontaneous symmetry breaking that follow are equally valid for an-harmonic
potentials. Let us first identify the collective dynamics which
describe the spontaneous symmetry breaking of this short-ranged microscopic
Hamiltonian.

In the standard treatment of the quantum crystal one begins by introducing new
coordinates, which are the displacements of atoms from their equilibrium
position. Then, after a Fourier transform the eigenstates of this Hamiltonian
are easily found. We take a slightly longer route by introducing bosonic
phonon operators from the very beginning and diagonalizing the quadratic part
of the Hamiltonian by performing a Bogoliubov transformation at the end. In
doing so we do not have to introduce any equilibrium position of the
atoms. Instead we can keep track of the center of mass motion of the crystal
as a whole, and this brings to the fore the thin spectrum in a natural
manner. Moreover we can use the exact same procedure in the next section to
find the collective order parameter dynamics for antiferromagnets.

The momentum and position operators are expressed in terms of bosonic operators as
\begin{eqnarray}{p}_j = i C \sqrt{\frac{\hbar}{2}} (b^{\dagger}_{j} - b^{\phantom\dagger}_j) ;
~~{x_j} = \frac{1}{C}\sqrt {\frac{\hbar}{2}} (b^{\dagger}_{j} + b^{\phantom\dagger}_j), 
\end{eqnarray}
so that the commutation relation  $[{x_j}, {p_{j'}}] = i \hbar \delta_{j,j'}$
is fulfilled. We choose $C^2 = \sqrt{2m \kappa}$ so that the Hamiltonian reduces to
\begin{eqnarray}
H = \frac{\hbar}{4} \sqrt \frac{2 \kappa}{m} \sum_{j,j'} 2 
(b^{\dagger}_{j} b^{\phantom\dagger}_j + b^{\phantom\dagger}_j b^{\dagger}_{j}) - (b^{\dagger}_{j} + b^{\phantom\dagger}_j) (b^{\dagger}_{j'}+b^{\phantom\dagger}_{j'})
\end{eqnarray}
and after a Fourier transformation 
\begin{eqnarray} H &=&  
\hbar \sqrt{\frac{\kappa}{2m}} \sum_k \left[ A_k b^{\dagger}_{k} b^{\phantom\dagger}_k + \frac{B_k}{2} 
(b^{\dagger}_k b^{\dagger}_{-k} + b^{\phantom\dagger}_k b^{\phantom\dagger}_{-k}) +1 \right], \nonumber 
\end{eqnarray}
where $A_k =2 - \cos \left( ka \right)$, $B_{k} = - \cos \left( ka \right)$
and $a$ is the lattice constant. This Hamiltonian is still not diagonal, since
the terms $b^{\dagger}_k b^{\dagger}_{-k}$ and $b^{\phantom\dagger}_{k}
b^{\phantom\dagger}_{-k}$ create and annihilate two bosons at the same
time. We get rid of these terms by a Bogoliubov transformation (see
appendix). After this the Hamiltonian in terms of transformed boson operators
$\beta^{\phantom\dagger}_k=\cosh(u_k) b^{\phantom\dagger}_{-k} + \sinh(u_k)
b^{\dagger}_k$ is
\begin{eqnarray}
H &=& \hbar \sqrt \frac{\kappa}{m} \sum_k \left[ 2 \sin |ka/2| \left( \beta^{\dagger}_{k}
\beta^{\phantom\dagger}_{k} + \frac{1}{2} \right) \right. \nonumber \\
&+& \left. \frac{1}{4} \sqrt {2}\cos\left(ka\right) \right]\nonumber \\ 
&=& 2 \hbar {\sqrt \frac{\kappa}{m}} \sum_k \sin |ka/2| \left[ n_k +\frac{1}{2} \right],
\label{eq:Hbeta}
\end{eqnarray}
since  $\sum_k \cos k = \frac{N}{2 \pi} \int^{\pi}_{- \pi} d k \cos
k = 0$. 

This result seems to coincide with the textbook Hamiltonian which we would
have obtained if we had followed the conventional route of Fourier
transforming the Hamiltonian for the displacements, and then quantizing
it. However, the Bogoliubov transformation has the advantage that it brings to
the fore a rather subtle point. When $k \to 0$ the excitation energy $\omega_k
\to 0$ and the two parameters in the Bogoliubov transformation diverge:
$\sinh(u_k) \to \infty$ and $\cosh(u_k) \to \infty$. Precisely at $k = 0$ the
canonical transformation is no longer well defined. We therefore should
investigate the bosonic terms in the Hamiltonian with  $k = 0$
separately. This zero momentum part of the Hamiltonian describes the obvious
fact that the crystal as a whole carries a kinetic energy associated with the
combined mass of all of its constituents, and is given by
\begin{eqnarray} H_{{\bf k} = 0} &=& \hbar \sqrt {\frac{\kappa}{2m}} \left( b^{\dagger}_{0} b^{\phantom\dagger}_0 - \frac{1}{2} \left(b^{\dagger}_{0} b^{\dagger}_{0} + b^{\phantom\dagger}_0 b^{\phantom\dagger}_0 \right)+1\right) \nonumber \\ 
& =& \hbar \sqrt{\frac{\kappa}{2m}} \left[ 1- \frac{1}{2} \left(b^{\dagger}_{0} - b^{\phantom\dagger}_0 \right)^2 \right]
\end{eqnarray}
where $(b^{\dagger}_{0} - b^{\phantom\dagger}_0)^2 = \frac{-2}{\hbar \sqrt {2 m \kappa}} {\bf p}^{*}_{0} {\bf p}^{\phantom *}_0$ so that
\begin{eqnarray}
H_{{\bf k}=0}= \frac{p^2_{tot}}{2 N m} + \text{constant},
\label{Hcoll}
\end{eqnarray}
where ${\bf p}_{tot} \equiv \sum_j {\bf p}_j=\sqrt{N} {\bf p}_{{\bf k}=0}$ is
the total momentum of the entire system, or equivalently, its center of mass
momentum. When $N$ is large, this Hamiltonian has states that are very low in
energy. These states in fact form the thin spectrum of the harmonic
crystal. We call this part of the spectrum {\it thin} because it contains so
few states of such low energy that its contribution to the free energy in the
thermodynamic limit completely disappears (see appendix). In turn, this implies that
these thin spectrum states do not contribute to any thermodynamically
measurable quantities such as for instance the specific heat of the crystal. Their
effect on the properties of the crystal is thus increasingly subtle, but its
existence can nonetheless have profound consequences. In classical systems
this thin spectrum is absent: it is quantum mechanics that generates it. About
a decade ago the deep meaning of the thin spectrum for interacting quantum
systems became clear and consequently its explicit mathematical structure was
determined~\cite{Kaplan90,Kaiser89}. 

%that we will discuss in more detail in the next section.
The groundstate of the Hamiltonian at ${\bf k}=0$, which governs the
collective behavior of the crystal as a whole, obviously has total momentum
zero. It thus has no uncertainty in total momentum and maximum uncertainty in
total position: translational symmetry is unbroken. Symmetry breaking can
occur if we add to the Hamiltonian of equation \eqref{Hcoll} a symmetry
breaking field of the form $B {\bf x}_{tot}^2/2$, where the center of mass
coordinate is ${\bf x}_{tot} \equiv \sum{_j} {\bf x}_j$. This yields a
harmonic oscillator equation for the collective position coordinate. Its well
known groundstate wavefunction is
\begin{eqnarray}
\psi_0 (x_{tot}) =\left( \frac{m \omega N}{\pi\hbar} \right)^{1/4} e^{-\frac{m\omega N}{2 \hbar} x_{tot}^2}; ~~\omega=\sqrt{\frac{B}{m}}.
\end{eqnarray}
This state describes a wave-packet for the center of mass coordinate in real
space, which of course corresponds to an equivalent superposition of total
momentum states: the symmetry breaking field $B$ couples the different thin
spectrum states of the crystal.
For a vanishing symmetry breaking field $B$ and finite number of atoms we have
$\omega N \rightarrow 0$ and the collective coordinate is completely
delocalized, as before: $\psi_0 (x_{tot})=const$. But taking the thermodynamic
limit ($N \rightarrow \infty$) {\it in presence of a finite symmetry breaking
  field} gives $\omega N \rightarrow \infty$ and the center of mass position
becomes completely localized in the center of the potential well ($\psi_0
(x_{tot})=\delta_{{\bf x}_{tot},0}$), even if at the end the symmetry breaking
field is sent to zero. As we already pointed out, such a singular limit
characterizes spontaneous symmetry breaking; in this particular case the
translational symmetry of the crystal as a whole is spontaneously broken. 
The occurrence of a thin spectrum which consists of the states
associated with the quantum mechanics of the macroscopic body as a whole is a
universal notion. Whenever a system exhibits a continuous symmetry
which is broken in the classically realized, macroscopic state, then
consequently there must be a spectrum of states associated
with the symmetry-restoring fluctuations of the orderparameter as a
whole. The universality of this notion is guaranteed by the Goldstone
protection which holds for the low energy excitations at $k \to 0$ in any system with a
broken continuous symmetry. The smallness of the energy spacing within the
thin spectrum warrants the orderparameter dynamics of macroscopic bodies to
take place on a time scale much larger than anything observable.

\subsection{Decoherence}
To study the effect of the thin spectrum on the coherence of many particle
qubits, let us first investigate the dynamics of such a qubit in the most general
terms. Consider a many particle system that is large enough to display a spontaneously broken
continuous symmetry, but small enough to be used as a qubit. This qubit will
then have a thin spectrum which we can label by the quantum number $n$. At the
same time the system must have two accessible quantum states that can be
used as the qubit states, and which can be labeled by the quantum number $m$.
Because we have no experimental control over the thin spectrum states, we will
have to start out the experiment with a thermal mixture of those states:
\begin{eqnarray}
\rho_{t<t_0} 
&=& \frac{1}{Z} \sum_{n} e^{-\beta E^{n}_{0}} \left|0,n\right> \left< 0,n\right|
\end{eqnarray}
where $\rho_t$ is the density matrix, $E^n_m$ is the energy of the state
$\left|m,n\right>$ and where, by definition, the partition function is $Z =
\sum_{n} e^{-\beta E_0^n}$ and $\beta^{-1}=k_BT$.
To begin using this qubit in a quantum computation we will typically have to
prepare it in some coherent superposition of the states in the two level
system. To do this we apply a rotation that takes the state $\left|0,n\right>$
into the state $\sqrt{1/2}\left( \left|0,n\right> + \left|1,n\right> \right)$
for all values of $n$. The resulting density matrix then is given by
\begin{eqnarray}
\rho_{t=t_0} 
&=& \frac{1}{2 Z} \sum_{n} e^{-\beta E^{n}_{0}}\left( |0,n\rangle\langle0,n| + |0,n\rangle\langle1,n| \right. 
\nonumber \\
&+& \left. |1,n\rangle\langle 0,n| + |1,n\rangle\langle1,n| \right). 
\label{eq:init}
\end{eqnarray}
If we know the Hamiltonian $H$ which governs the dynamics of the qubit, then we can follow the
time evolution of this density operator by applying the time evolution
operator $U |m, n\rangle \equiv e^{-\frac{i}{\hbar} H(t-t_0)} |m, n\rangle = e^{-\frac{i}{\hbar}
  E^n_m (t-t_0)} |m,n\rangle$. We then find for the density matrix at $t>t_0$
\begin{eqnarray}
\rho_{t>t_0} &=& U \rho_{t=t_0} U^{\dagger} \nonumber\\
&=& \frac{1}{2 Z} \sum_{n} e^{-\beta E^{n}_{0}} \left( |0,n\rangle\langle0,n| + |1,n\rangle\langle1,n|  \right. \nonumber \\
&+&
\left. \left[e^{-\frac{i}{\hbar}(E^{n}_{0}-E^{n}_{1}) (t-t_0)} |0,n\rangle\langle1,n| +  h.c. \right] 
\right),
\label{eq:one}
\end{eqnarray}
where $h.c.$ denotes the hermitian conjugate of its preceding term.

Experimentally the thin spectrum is as good as unobservable because of its
extremely low energy and its vanishing thermodynamic weight (see appendix). We
therefore have to trace these states out of the density matrix. This will
yield a reduced (observable) density matrix, defined by
\begin{eqnarray}
\rho^{red}_{t>t_0} = \sum_{j} \langle j|\rho_{t>t_0}|j\rangle
\label{eq:two}
\end{eqnarray}
where the trace is over thin spectrum states labeled by $j$ and $\langle
j|m,n\rangle \equiv |m\rangle \delta_{j,n}$. Performing the trace, we find the
following reduced density matrix in the basis of the states $\left|m\right>$
\begin{eqnarray}
\rho^{red}_{t>t_0}=\frac{1}{2} 
\left [ 
\begin{array}{cc} 1 & \rho^{OD}_{t>t_0} \\  \left[\rho^{OD}_{t>t_0}\right]^{\rm *} & 1 \end{array} 
\right ] ,
\label{eq:three}
\end{eqnarray}
where the off-diagonal matrix element is defined as
\begin{eqnarray}
\rho^{OD}_{t>t_0} \equiv \frac
{1}{Z} \sum_{n} e^{-\beta E^{n}_{0}} e^{-\frac{i}{\hbar}(E^{n}_{0}-E^{n}_{1})
  (t-t_0)}.
\label{elementOD}
\end{eqnarray}
If this off-diagonal matrix element vanishes at some time, then
the qubit will have decohered at that time, due to the presence of the thin
spectrum. In general $\Delta E_{thin} \equiv E^{n}_{0} - E^{n}_{1}$ will not
be zero, and this shift in energy corresponds to a phase shift of the thin
spectrum states. These phases will typically interfere destructively,
lowering $\rho^{OD}_{t>t_0}$ and leading to dephasing and decoherence.
The timescale for this decoherence process is set by the inverse of the
involved energy scale, and will therefore be proportional to $\hbar / \Delta
E_{thin}$.

For this dephasing to occur however, it is necessary that a finite number of
thin spectrum states participates in the dynamics of decoherence. How many
states do contribute to the process is governed by the Boltzmann factor
$e^{-\beta E^{n}_{0}}$, which exponentially suppresses states of energy higher
than $\sim k_B T / E_{thin}$ with $E_{thin}$ the typical level spacing of the
thin spectrum states. Putting these arguments together, one finds that the
characteristic timescale on which $\rho^{OD}_{t>t_0}$ will vanish should be
proportional to
\begin{eqnarray}
t_{spon} \propto \frac{\hbar}{k_B T} \frac{E_{thin}}{\Delta E_{thin}}.
\end{eqnarray}

In the following sections we will calculate $t_{spon}$ explicitly for a number
of realizations of the many particle qubit. We will see that in the generic
situation $\Delta E_{thin} \propto E_{thin} / N$ so that we find
\begin{eqnarray}
t_{spon} \propto \frac{N \hbar}{k_B T},
\end{eqnarray}
which is our main result. 

\section{The Crystal as a Qubit}
As a first example of the influence of the thin spectrum on coherence, let us
try to employ the harmonic crystal discussed in the introduction as a
qubit. In order to do so we will have to define a set of two states that are
to be used as the calculational states of the qubit. A simple choice for such
a set could be to use the presence or absence of an interstitial
excitation. This leads to the definition of the state $\left|m=0\right>$ describing the crystal
with $N$ atoms, and the state $\left|m=1\right>$ which has one extra
interstitial atom, and is described by the same model, but with $N+1$ atoms in
the lattice. The thin spectrum is exactly as described in~\eqref{Hcoll}, so
that the energy can be defined as
\begin{eqnarray}
E^n_m = \frac{n^2}{2M \left(N+m\right)} + \mu m,
\end{eqnarray}
where $M$ is the mass of an atom, $n$ labels states with different total
momentum (which make up the thin spectrum), and $\mu$ is the chemical
potential associated with adding an extra atom to the lattice.

We are now in the position to simply substitute this information into the
general expression for the off-diagonal matrix element of the reduced density
matrix~\eqref{elementOD}, yielding
\begin{eqnarray}
\rho^{OD}_{t>t_0} = & \frac{1}{Z} & e^{-\frac{i}{\hbar}\mu(t-t_0)} \sum_{n} e^{-\beta \frac{n^2}{2MN}}
\cdot \nonumber \\
&& e^{-\frac{i}{\hbar}\frac{n^2}{2M}\left(\frac{1}{N}-\frac{1}{N+1}\right)
  (t-t_0)}.
\end{eqnarray}
The constant phase factor $e^{-\frac{i}{\hbar}\mu(t-t_0)}$ does not contribute
to the decoherence process, but the terms depending on $n$ introduce phase
shifts into the dynamics of the system, which lead to the disappearance of
$\rho^{OD}_{t>t_0}$ over time. Upon introduction of $E_{thin}=1/(2MN)$ and $\Delta E_{thin}=1/(2MN(N+1)) \simeq
1/(2MN^2)$, a straightforward evaluation of the sum over thin spectrum states
yields $t_{spon}$, defined as the half time for $\left|\rho^{OD}_{t>t_0}\right|$
\begin{eqnarray}
t_{spon} &=& \frac{2 \pi \hbar}{k_B T} \frac{E_{thin}}{\Delta E_{thin}} \nonumber \\
&=& N \frac{2 \pi \hbar}{k_B T}.
\end{eqnarray}

By using the crystal as a qubit in this way we have assumed that we can just
ignore the symmetry breaking field as soon as the crystal has been localized
in space at some time in the past. In general this may not be true, because
the thermodynamic limit and the limit of disappearing localization field do
not commute. We should therefore also consider the situation in the presence
of a small but finite symmetry breaking field $B$. In that case the energies
of the system will be given by
\begin{eqnarray}
E^n_m = n \sqrt{\frac{B}{2M \left(N+m\right)}} + \mu m,
\end{eqnarray}
which will again lead to a phase factor which is constant in $n$, and a sum
over phases which can be written as multiples of
$E_{thin}=\sqrt{{B}/{2M\left(N+m\right)}}$ and $\Delta
E_{thin}=E_{thin}/N$. The summation over thin spectrum states will thus again
yield the coherence time
%
%\begin{eqnarray}
$t_{spon} = N \frac{2 \pi \hbar}{k_B T}.$
%\end{eqnarray}
%

\subsection{Goldstone Modes}
The interstitial excitation used in the previous section to characterize the qubit
state of the harmonic crystal is a very rough excitation to use for that
purpose. The extra atom in the crystal will immediately affect the whole
lattice structure, and thus couple to many phonon-excitations. On top of that
it also automatically increases the mass of the crystal as a whole, and
thereby influences the thin spectrum directly. Because of this direct
alteration of the crystal properties, serious decoherence effects are to be
expected. In constructing a qubit using the quantum crystal it is therefore
better to look for a more 'silent' excitation. These silent excitations are
naturally found in the long wavelength Goldstone modes of the crystal,
i.e. the low energy phonons.

To see what is the effect on the thin spectrum of having phonons present,
let's consider the general anharmonic term
\begin{eqnarray}
H_{an}=\kappa_z \sum_j \left( {\bf x}_j - {\bf x}_{j+1} \right)^z
\end{eqnarray}
which supplements the Hamiltonian~\eqref{eq:Xtal}, and introduces
phonon-phonon interactions in its dynamics. The Fourier transform of $H_{an}$
is 
\begin{eqnarray}
H_{an}=\kappa_z \sum_j \left[ \frac{1}{\sqrt{N}}\sum_k e^{i k j} \left( 1 -
  e^{i k a} \right) {\bf x}_k \right]^z
\end{eqnarray}
with $a$ the lattice constant. From this expression it is immediately
clear that for all values of $z$ the ${\bf k}=0$ part of $H_{an}$ will
vanish. This is of course just a manifestation of the Goldstone protection of
the long wavelength phonon modes, but it implies that the presence of phonons can not alter
the form of the thin spectrum part of the Hamiltonian given
by~\eqref{Hcoll}. The addition of a symmetry breaking field will not modify
that conclusion, because the energies of the collective model in presence of a
localization field only depend on the total mass of the crystal and the size
of the symmetry breaking field. The phonons do not influence the mass of the quantum
crystal, and thus we find that $\Delta E_{thin} = 0$ both for zero and finite
symmetry breaking field $B$.

Because the phonons do not influence the thin spectrum at all, there will also
be no induced decoherence of a superposition of different phonon states by the
thin spectrum. In this case thus
%
%\begin{eqnarray}
$t_{spon} \to \infty.$
%\end{eqnarray}
%
Of course the superposition of phonon states will be influenced by all the
other sources of decoherence that are present in the crystal, but it is
protected from the decohering effects of the thin spectrum by the fact that
phonons do not influence the properties of the crystal as a whole.

The fact that the Goldstone modes of the crystal supply an excitation that is
protected from all decohering effects of the thin spectrum 
%may be due to the
%simple Abelian nature of the model. 
is due to the fact that we only considered translation symmetry breaking of the 
crystal {\it as a whole}. Within it, there is still translation invariance of the atoms. This pathology that is due to the periodic boundary conditations that we implicitely assumed. A coupling between the thin spectrum and the Goldstone modes can appear as soon as this internal symmetry is broken as well --which is the generic situation.
In fact we will show in the next section
that even in the most silent non-Abelian model that one can think of, i.e. the
antiferromagnetic Lieb-Mattis model, this pathology will disappear.

\section{Lieb-Mattis model}
Let us now turn to the discussion of the antiferromagnetic Lieb-Mattis model.
The reason for considering the rather particular, long ranged Lieb-Mattis
model is that for a broad class of Heisenberg models with short-ranged
interactions it constitutes the effective Hamiltonian for the thin
states. Similar collective models underly the breaking of other continuous
symmetries, such as for instance gauge symmetry in a superconductor. In that
case the collective Hamiltonian turns out to be very similar to the
Lieb-Mattis Hamiltonian as far as the structure of the thin spectrum and the
composition of the wavefunction of the symmetry broken state are concerned. To
explicitly show how the Lieb-Mattis model is arises from a Heisenberg model,
let us consider an antiferromagnet on a bipartite lattice with isotropic
nearest neighbor interactions between quantum spins of size $\sigma$. Its
Hamiltonian is 
\begin{eqnarray}
H = J \sum_{i, \delta} {\bf S}_i {\bf S}_{i+\delta},
\label{eq:H0}
\end{eqnarray}
where $i$ labels all the spins on the $A$ sublattice, and the $\delta$ are the
vectors connecting site $i$ to its neighbors on sublattice B. The
generalization to other types of interactions and even other types of lattices
is straightforward~\cite{Bernu92, Bernu94, Capriotti01, Ziman, Fisher, Gross}. 
The magnon spectrum of this Hamiltonian can be found within linear spin wave
theory. One approximates the spin operators with Holstein-Primakoff bosons as
\begin{eqnarray}
S_{i \epsilon A}^z \to& \sigma-a^{\dagger}_i a^{\phantom\dagger}_i,& S_{i \epsilon B}^z \to b^{\dagger}_i b^{\phantom\dagger}_i -\sigma, \nonumber \\
S_{i \epsilon A}^+ \to& \sqrt{2 \sigma} a^{\phantom\dagger}_i,& S_{i \epsilon B}^+ \to \sqrt{2 \sigma} b^{\dagger}_i, \nonumber \\
S_{i \epsilon A}^- \to& \sqrt{2 \sigma} a^{\dagger}_i,& S_{i \epsilon B}^- \to \sqrt{2 \sigma} b^{\phantom\dagger}_i.
\end{eqnarray}
To quadratic order in the boson operators the Hamiltonian becomes, after a Fourier transformation 
\begin{eqnarray}
H^{LSW} = \frac{1}{2}J N z \sigma^2 &+& J z \sigma \sum_{\bf k} \left( (a^{\dagger}_{\bf k} a^{\phantom\dagger}_{\bf k} + b^{\dagger}_{\bf k} b^{\phantom\dagger}_{\bf k}) \right. \nonumber \\
&+& \left. \gamma^{\phantom\dagger}_{\bf k} (a^{\dagger}_{\bf k} b^{\dagger}_{\bf -k} + a^{\phantom\dagger}_{\bf k} b^{\phantom\dagger}_{\bf -k}) \right),
\end{eqnarray}
where $z$ is the coordination number of the lattice, $N$ the number of lattice sites and $\gamma^{\phantom\dagger}_k \equiv \frac{1}{z} \sum_{\delta} e^{i k \delta}$. The last two terms in this expression can be diagonalized by a Bogoliubov transformation (see appendix).
Again the important point is that the Bogoliubov transformation is singular at $k=0$ and $k=\pi$, as in both cases $\gamma_k^2 \to 1$. We therefore treat these two $k$-points separately. Turning back to the notation in terms of spins, using that the Fourier transform of our Hamiltonian is
\begin{eqnarray}
H = J \sum_{\bf k} \gamma_{\bf k} {\bf S}_{\bf k} \cdot {\bf S}_{-{\bf k}}
\end{eqnarray}
and
\begin{eqnarray}
{\bf S}_{{\bf k}={\bf 0}} &=&  \frac{1}{{\sqrt N}} \sum_{i \epsilon A,B} {\bf S}_i = \frac{1}{{\sqrt N}} ({\bf S}_A+{\bf S}_B), \nonumber \\
{\bf S}_{{\bf k}={\bf \pi}} &=& \frac{1}{{\sqrt N}} ({\bf S}_A-{\bf S}_B) \nonumber
\end{eqnarray}
we find that the singular parts of the spectrum reduce exactly to the Lieb-Mattis Hamiltonian
\begin{eqnarray}
H &=& H^{sym}_{LM} + J \sum_{{\bf k} \neq {\bf 0},{\bf \pi}} \gamma_{\bf k}
{\bf S}_{\bf k} \cdot {\bf S}_{-{\bf k}} \nonumber \\
H^{sym}_{LM} &=& \frac{2J}{N} {\bf S}_A \cdot {\bf S}_B = \frac{J}{N}({\bf S}^2 - {\bf S}^2_A - {\bf S}^2_B),
\label{HLM}
\end{eqnarray}
where ${\bf S}_A$ and ${\bf S}_B$ are the total spins of each sublattice, and
${\bf S}$ is the {\em total} spin of the system. 
From here on we will focus entirely on this collective Hamiltonian, as it is
the only part of the Heisenberg-like Hamiltonians that is relevant for the
spontaneous symmetry breaking of the antiferromagnet as a whole. Notice that
the {\it internal} ordering of the individual spins within the
antiferromagnet can be destroyed by fluctuations of finite wavelength that we do not consider in
this collective, long ranged model. 
We assign to the Hamiltonian $H^{sym}_{LM}$ the superscript $sym$ because this Hamiltonian, as we will show below, describes the symmetric (symmetry unbroken) state of the antiferromagnet.
%Note that in a $d$-dimensional system the number of spins is $N=L^d$, where $L$ is the linear extent of the system. So the energy scale for the lowest possible spin wave (magnon) excitation is $J/L$. However the energy scale of the thin spectrum is $J/N=J/L^d$, and is thus --in any dimension higher than one-- much lower than the magnon energy scale.
% So the Lieb-Mattis Hamiltonian describes collective excitations of the system, the thin states, at extremely low energy.

In $H^{sym}_{LM}$ each spin on the $A$ sublattice interacts with all spins on
the $B$ sublattice and vice versa, thus creating infinite range
interactions. The energies of the Hamiltonian are trivially identified as $
\frac{J}{N}[S(S+1)-S_A(S_A+1)-S_B(S_B+1)]$ and the corresponding
eigenfunctions are labeled by their quantum numbers $|S_A, S_B, S,
M\rangle$. Here the $z$-component of the total spin ${\bf S}$ is denoted by
$M$. Clearly the ground state of $H^{sym}_{LM}$ is a singlet of total spin:
the state with lowest energy has $S=0$. In fact there is an exact proof that
the groundstate of any finite spin system of this sort is a total spin zero
($S=0$) singlet~\cite{Auerbach94}. Notice that all states which differ only in
$M$ are degenerate. For simplicity (and without loss of generality) we
henceforth take the quantum number $M$ to be zero \cite{Lieb62}.

The groundstate singlet $|N\sigma/2, N\sigma/2, 0, 0\rangle$, with both $S_A$
and $S_B$ maximal and $S=M=0$ is separated by energies of order $J/N$ from
states with higher $S$. The set of these extremely low energy states that only
differ in their total spin quantum number forms the thin
spectrum~\cite{Anderson52,Lieb62,Kaplan90}. Since~\eqref{HLM} is contained
in~\eqref{eq:H0} as its ${\bf k}={\bf 0}$ and ${\bf k}={\bf \pi}$ components,
and since the thin spectrum of the Lieb-Mattis model is formed by the ${\bf
  k}={\bf 0}$ component, exactly the same thin spectrum must govern the
collective dynamics of other antiferromagnets with short-range
interactions~\cite{Bernu92, Ziman, Fisher}.

There are also excitations in~\eqref{HLM} that can be created by lowering $S_A$ or $S_B$. This costs an energy of order $J$, and it can easily be shown that these excitations correspond to the elementary excitations, the magnons, of the Lieb-Mattis system~\cite{Mattis}. Because of the extremely long ranged interactions the magnons are gapped and dispersionless. 

\subsection{The Lieb-Mattis Hamiltonian with symmetry breaking field} 
Having defined the Lieb-Mattis model in its symmetric form, we now review how
to explicitly break its non-Abelian SU(2) spin rotation symmetry. We will show
that in the thermodynamic limit the symmetry breaking occurs
spontaneously~\cite{Anderson52}. Since the groundstate of $H^{sym}_{LM}$ is a
singlet of total spin, this state is orthogonal to the N\'eel state, which is
the ground state of a classical antiferromagnet. We should stress here that
there is a marked difference between ferro- and antiferromagnets. Even if
spontaneous symmetry breaking is very often discussed with the example of a
ferromagnet at hand, the spontaneous symmetry breaking in a ferromagnet is not
the generic situation for an interacting quantum system. The reason is that
the total magnetization (pointing along, e.g., the z-axis), which is the order
parameter of a ferromagnet, commutes with the Hamiltonian: it is nothing but
the projection of the total spin along that axis, $S^z_{tot}$. So the
situation arises that the orderparameter is a constant of motion, which is a
pathology of the ferromagnet. This same pathology leads to the absence of an
interesting thin spectrum, because in the ferromagnet states with different
$S^z_{tot}$ are strictly degenerate. Quantum systems in general, however, have
non-trivial thin spectra.

Refocusing on antiferromagnets, we need to proof that the N\'eel state is a
stable groundstate in the thermodynamic limit. In order to do so, an explicit
symmetry breaking field $B$ is introduced~\cite{Kaiser89}:
\begin{eqnarray}
H_{LM} = H^{sym}_{LM} - B (S^z_A - S^z_B).
\end{eqnarray}
Clearly the symmetry breaking field induces a finite sublattice magnetization. The field couples the different total spin states of the thin spectrum by the matrix elements
\begin{widetext}
\begin{eqnarray}
\langle S_A, S_B, S, M| S^z_A - S^z_B|{S'}_A, {S'}_B, {S'}, {M'} \rangle = 
\delta_{S_A,{S'}_A} \delta_{S_B,{S'}_B} \delta_{M,{M'}} 
\left[ f_{S+1} \delta_{S,{S'}-1} + g_S  \delta_{S,{S'}} + f_S  \delta_{S,{S'}+1} \right]
\end{eqnarray}
where
\begin{eqnarray}
f_S  \equiv \sqrt{\frac{ \left[S^2-(S_A-S_B)^2\right]\left[(S_A+S_B+1)^2-S^2\right]\left[S^2-M^2\right]  }
{ (2 S+1)(2 S -1)S^2  }}
\ \ {\rm and} \ \  
g_S  \equiv \frac{(S_A-S_B)(S_A+S_B+1)M}{S(S+1)}.
\end{eqnarray}
\end{widetext}
These matrix elements are found by performing a rather tedious sum over Clebsch Gordon coefficients in the following expression (see also the appendix):
\begin{eqnarray}
&& \langle S_A, S_B, S, M| S^z_A - S^z_B |{S'}_A, {S'}_B, {S'}, {M'} \rangle  = \nonumber \\
&&\sum_{M_A}  \left[ C^{S,M}_{S_A,S_B,M_A,M-M_A} C^{{S'},M}_{S_A,S_B,M_A,M-M_A}  
(2 M_A - M) \right] \nonumber \\
&& \ \ \ \ \ \ \delta_{S_A,{S'}_A} \delta_{S_B,{S'}_B} \delta_{M,{M'}}.
\end{eqnarray}
The spectrum of eigenstates $|n\rangle$ in the presence of a symmetry breaking field can now be found by expanding these states in the basis of total spin states: $|n\rangle=\sum_{S} u^n_S |S\rangle$ (for clarity of notation we suppress the dependency of $u^n_S$ and other variables on the quantum numbers $S_A$, $S_B$ and $M$). In this basis, Schr\"{o}dinger's equation becomes~\cite{Kaiser89}
\begin{eqnarray}
&H_{LM} |n\rangle = E^{n}_{0} |n\rangle  \Leftrightarrow \nonumber \\
&\sum_{S} \left[  \frac{JS(S+1)}{N} u^n_S + E^{sym}_{LM} u^n_S  - B f_{S+1}  u^n_{S+1} \right. \nonumber \\
&\left. - B f_S  u^n_{S-1}  \right] |S \rangle = E^{n}_{0} \sum_{S}  u^n_S ~|S\rangle,
\label{eq:SE}
\end{eqnarray}
where $E^{sym}_{LM}$ is the groundstate energy of $H^{sym}_{LM}$ and $E^{n}_{0}$ is the energy of eigenstate $|n\rangle$ of the {\it symmetry broken} Hamiltonian --its thin spectrum.  Here we restricted ourselves to the zero-magnon subspace, where $S_A=S_B=N \sigma /2$ (hence the subscript $0$ in $E^{n}_{0}$). The generalization to systems with a finite number of magnons will be straightforward.

In the continuum limit where $N$ is large and $0 \ll S \ll N$, the matrix elements due to the symmetry breaking field simplify considerably. It is easy to show that in this case 
\begin{eqnarray}
f_S  \simeq \frac{N\sigma}{2}\sqrt{1-\left(\frac{S}{N\sigma}\right)^2} \simeq N\sigma/2.
\end{eqnarray} 
We will see shortly that only the first $\approx \sqrt{N}$ total spin states
contribute to the groundstate wavefunction, so that an expansion in $S/N$ is
justified. Notice that when the sublattice spin $S_A$ is reduced by one,
i.e. when there is a spin-wave present, the matrix element $f_S$ is reduced:
$f_S^1 \approx f_S \frac{N\sigma-1}{N\sigma} = f_S \left( 1-\frac{1}{N\sigma}
\right)$, for large $N$. This reflects the fact that a magnon reduces the
N\'eel order parameter (the staggered magnetization) by unity. This effect is
small, but turns out to be essential when we shall consider the quantum
coherence of magnons: dephasing will occur because magnons give rise to a
subtle change in the level splitting of the thin spectrum. This change in level splitting turns out to be inversely proportional to $N$, the total number of spins in the antiferromagnet.

\begin{figure}
\includegraphics[width=0.49\columnwidth]{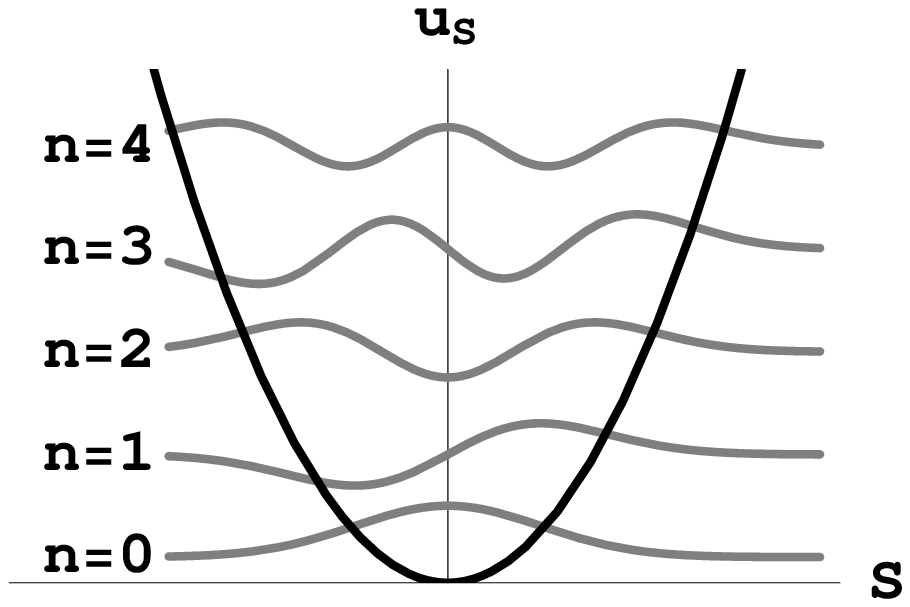}
\includegraphics[width=0.49\columnwidth]{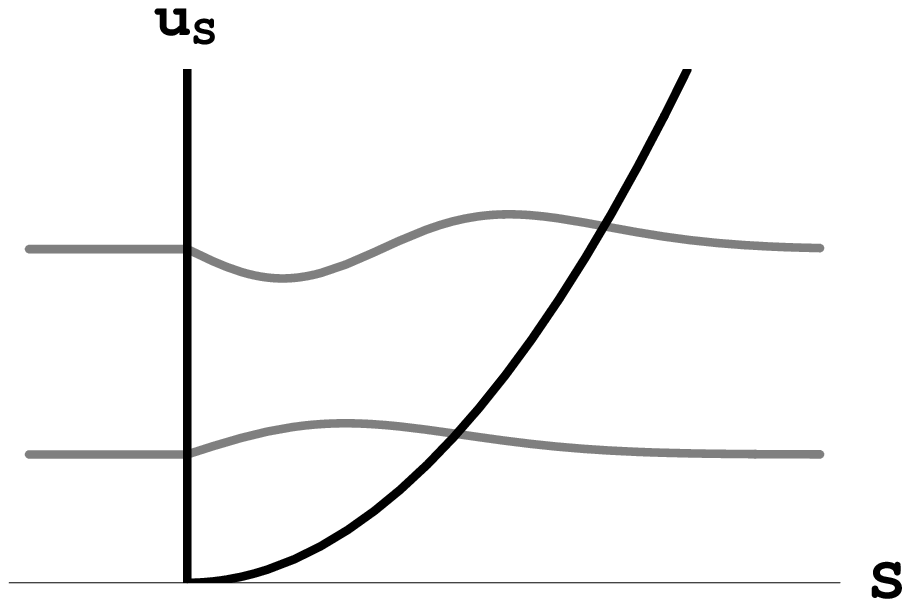}
\caption{Wavefunctions of the thin-spectrum state in presence of a symmetry
  breaking field, in the continuum limit. The boundary condition $S \ge 0$ implies that of the harmonic oscillator solutions (left) only the odd ones are allowed (right), as these have a node at the origin.} 
\label{fig:harmonic}
\end{figure}

\begin{figure}
\includegraphics[width=0.8\columnwidth]{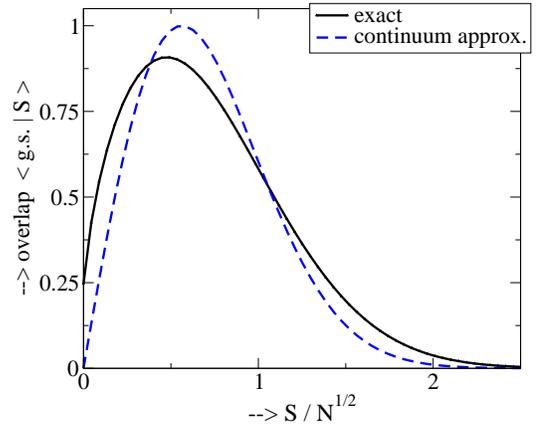}
\caption{Comparison of the exact symmetry broken N\'eel wavefunction (for
  $N=500$ spins) and the N\'eel wavefunction in the continuum limit (a
  harmonic oscillator eigenstate). The overlap of the N\'eel state with the
  different total spin states is shown as a function of the total spin quantum
  number. The parameters are $J=1$ and the symmetry breaking field is $B=1/10$. The wavefunctions are rescaled such that the maximum of the harminic oscillator wavefunction is unity.}
\label{fig:wave}
\end{figure}

In the continuum limit the Schr\"{o}dinger's equation~\eqref{eq:SE} reduces to
\begin{eqnarray}
-\frac{1}{2} \frac{\partial^2}{\partial S^2} u^n_S + \frac{1}{2} \omega^2 S^2 u^n_S = \nu_n u^n_S,
\label{SchEq}
\end{eqnarray}
where again we have used $0 \ll S \ll N$. In this equation we introduced
$\omega=\frac{1}{N} \sqrt{\frac{2J}{B\sigma}}$ and
$\nu_n=\frac{E^n_0-E^0_{LM}}{B N \sigma} + 1$. Obviously this is the
differential equation of a harmonic oscillator. The eigenstates $u^n_S$ thus
are well known and the corresponding eigenvalues are $\nu_n = (n+1/2)\omega$, so that
\begin{eqnarray}
E^{n}_{0} = E^{sym}_{LM} - B N \sigma + \left( n+\frac{1}{2} \right) E_{thin},
\label{eq:En0}
\end{eqnarray}
where the quantum of energy for the states labeled by $n$ is $E_{thin}=\sqrt{2
  \sigma J B}$. For the harmonic oscillator $n$ is a non-negative
integer. However, in the present situation we have to meet the boundary
condition that $S \geq 0$ or, equivalently, that $u^n_S = 0$ if $S < 0$. So
$u^n_S$ has to vanish at the origin. This boundary condition is trivially met
by eigenfunctions that are odd and have a node at $S=0$, see
Fig.~\ref{fig:harmonic}. Thus solutions to the Schr\"{o}dinger's
equation~\eqref{SchEq}  are harmonic oscillator eigenfunctions of order $n$,
where $n$ is an odd positive integer. In Fig.~\ref{fig:wave} the groundstate wavefunction in the continuum limit is compared with the exact wavefunction for large $N$. It makes clear that the continuum approximation is very good one.

Let us consider the energy spectrum in Eq.~\eqref{eq:En0} in more detail.
%First, we observe that the zero-point energy associated with global fluctuations of the order parameter equal to $\frac{3}{2}E_{thin}$. 
Clearly if $B$ is zero we recover the groundstate energy $E^{sym}_{LM}$ of the
symmetric case that we discussed before. However, if there is a finite
staggered field $B$, there is a gain in groundstate energy proportional to
$BN$, which reveals that the energy spectrum in Eq.~\eqref{eq:En0} is the one
of a N\'eel state. The same conclusion is reached by directly calculating the
orderparameter (see appendix). The result is shown in
Fig.~\ref{fig:orderparameter}. Apparently, for the symmetry broken N\'eel
state to be stable, the symmetry breaking field can be exceedingly small, as
long as $N$ is large enough. In other words: in the thermodynamic limit the
spin rotation symmetry of $H^{sym}_{LM}$ can be {\it spontaneously} broken by
an infinitesimal field $B$. Putting it in a more formal manner: spontaneous symmetry breaking gives rise to the singular limit
\begin{eqnarray}
\lim_{N \to \infty} \lim_{B \to 0} \left\langle   \frac{S_A^z-S_B^z}{N \sigma}  \right\rangle &=& 0 {\rm ~~~and~~~} \nonumber \\ 
\lim_{B \downarrow 0} \lim_{N \to \infty} \left\langle   \frac{S_A^z-S_B^z}{N \sigma}  \right\rangle &=& 1.
\end{eqnarray}
This in fact defines spontaneous symmetry breaking, just as it did in the case
of the quantum crystal. That for the Lieb-Mattis Hamiltonian this limit is singular is directly clear from Fig.~\ref{fig:orderparameter}.

\begin{figure}
\includegraphics[width=0.8\columnwidth]{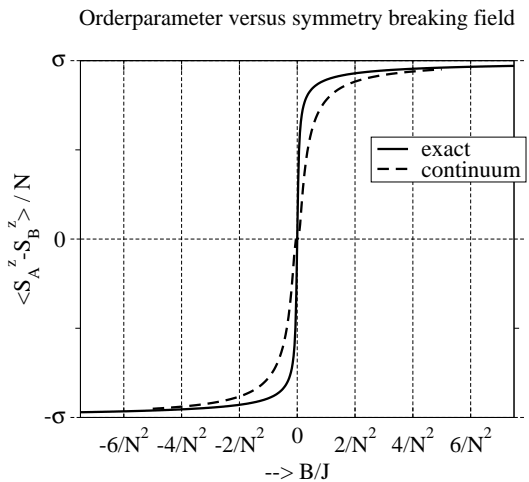}
\caption{Orderparameter as a function of symmetry breaking field. The exact
  result for $N=100$ spins, the continuum expression for the orderparameter is derived in the appendix.} 
\label{fig:orderparameter}
\end{figure}

In the symmetry broken N\'eel state the excitations labeled by $n$ act as a new thin spectrum with excitation energies that are multiples of $E_{thin}=\sqrt{2 \sigma J B}$. The magnon excitation energy is still of order $J$. 

\begin{figure}
\includegraphics[width=0.6\columnwidth]{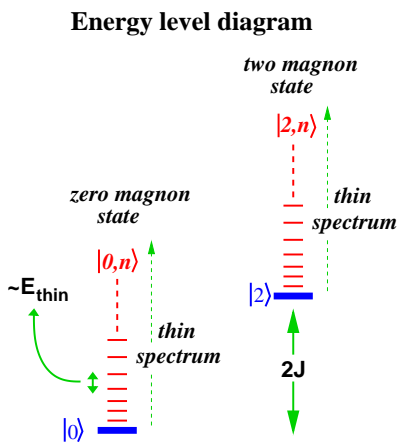}
\caption{Energy level scheme with the zero and two magnon states, each with its tower of thin spectrum states. The level spacing in the thin spectrum is $E_{thin}$, magnons live on an energy scale $J$.} 
\label{fig:thin}
\end{figure}

We now repeat the analysis above for a N\'eel state with $m$ magnons (by setting $S_A = S_B = N \sigma/2 - m/2$ and using $f^m_S$ instead of $f_S$). In this case the energy spectrum becomes
\begin{eqnarray}
E^n_m = E_0^n + m(2 \sigma J + B) - \frac{m}{2 N \sigma} (n+1/2) E_{thin},
\label{eq:Enm}
\end{eqnarray} 
see Fig.~\ref{fig:thin}. Note that as we stated before, there is a subtle effect of the magnons on the thin states: $m$ magnons cause a change in energy of the thin spectrum of the order of $m/N$. This effect will turn out to be essential for the decoherence mechanism that we will discuss in the following section. 

Physically the change in energy of the thin states due to the presence of a
magnon can easily be understood. If there are $m$ magnons present in the
antiferromagnet, then the order parameter of the total system is reduced by
$m$. Since the thin spectrum describes the global excitations of the order
parameter, its energy is proportional to the order parameter itself. The ratio
of the N\'eel order parameter of the excited state with $m$ magnons and the
one of the groundstate with a fully developed order parameter is
$(N-m)/N$. Therefore, when there are $m$ magnons present, the relative change
of the orderparameter is $m/N$ and the change in energy of the thin states is
therefore of the order $m E_{thin}/N$ which explains the last term in expression~(\ref{eq:Enm}).

\subsection{Preparing the many-spin qubit} 
Using the many-body Lieb-Mattis model
with $N$ spins and $\sigma=1/2$, we now study the coherence of the antiferromagnet when it is used as a
qubit. Again there are many ways in which one can define a two-level system to
be used as the qubit states. The best possible choice in this case is provided
by the gapped and dispersionless magnons: we use as a qubit (or cat-state) the
superposition of a perfectly ordered antiferromagnet and the state of the
antiferromagnet with one magnon on each sublattice. Due to the long-range
nature of the interaction in the Lieb-Mattis model the gapped magnons
themselves are not damped and as such do not decay or decohere. Also, in
analogy to the quantum crystal we expect the magnons or Goldstone modes of the
antiferromagnet to influence the thin spectrum as little as possible. A magnon
has an energy $J$, which we assume is the energy scale that is available to
the (thought-)experimentalist to prepare, manipulate and read out the
qubit~\footnote{Because of a technicality we choose to use a state with two
  magnons instead of a single one.  By doing so we ensure that our system
  stays in the subspace of zero total magnetization ($M=0$), which
  considerably simplifies the calculations without loss of generality.}.

\begin{figure}
\includegraphics[width=\columnwidth]{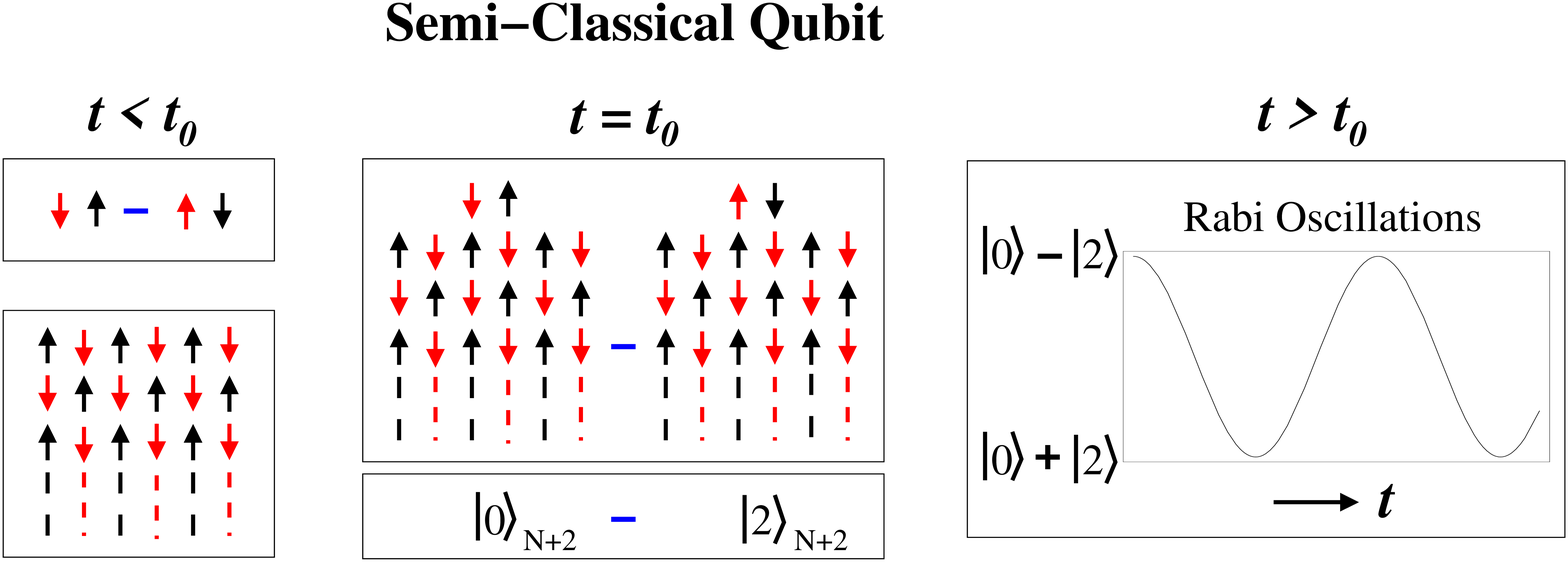}
\caption{Semi-classical time evolution of a two spin qubit that at $t=t_0$
  starts interacting with a $N$-spin Lieb-Mattis magnet, thus forming at
  $t>t_0$ a many-body qubit made out of $N+2$ spin. Quantum coherence is
  preserved at all times, since in this semi-classical approximation the thin
  spectrum is neglected.} 
\label{fig:qubit}
\end{figure}

With the exact expressions for all eigenstates and energies of both the symmetric and the symmetry broken Hamiltonian at hand, we are in the position to set up the initial state for our many-particle qubit. Instead of simply assuming that we are in a previously prepared superposition of states with zero and two magnons, we will explicitly construct this initial state. This can be done by coupling at time $t=t_0$ a two spin singlet to the symmetry broken $N$-spin Lieb-Mattis system, see Fig.~\ref{fig:qubit}. From this the desired superposition results.  So for times $t< t_0$, the Lieb-Mattis antiferromagnet is completely decoupled from the two spin singlet and the total wavefunction is thus the direct product of the wavefunctions of the $N$-spin magnet and the two-spin singlet state:
\begin{eqnarray}
|\psi_{t<t_0}\rangle = |0, n \rangle \otimes \left|singlet\right\rangle.
\end{eqnarray}
Here we denote the N\'eel state with $m$ magnons and $n$ thin spectrum excitations by $|m,n\rangle$. The state $|singlet\rangle$ is $\frac{1}{\sqrt{2}} \left[ | \downarrow_1 \uparrow_2 \rangle - | \uparrow_1 \downarrow_2 \rangle \right]$. Upon instantaneous inclusion at $t=t_0$ of the two spin state in the Lieb-Mattis lattice, the groundstate of the decoupled system at $t<t_0$ can be expressed in terms of the eigenstates of the $N+2$ spin system at $t=t_0$. The exact groundstate wavefunction is then given by the following formidable expression
\begin{widetext}
\begin{eqnarray}
|\psi_{t<t_0}\rangle = \sum_{S=0}^{N-1} u^n_S \left|S, 0\right\rangle \otimes \left|singlet\right\rangle 
&=& \sum_{S=0}^{N-1} u^n_S \sum C^{S,0}_{N/4,N/4,M_A,-M_A}  C^{0,0}_{1/2,1/2,M_1,-M_1} C^{S_{A1},M_A+M_1}_{N/4,1/2,M_A,M_1} \nonumber \\
&& C^{S_{B2},-M_A-M_1}_{N/4,1/2,-M_A,-M_1} C^{S_,0}_{S_{A1},S_{B2},M_A+M_1,-M_A-M_1} \delta_{S, S_{T}} |S_{A1}, S_{B2}, S, 0\rangle, 
\end{eqnarray}
\end{widetext}
where we sum the Clebsch Gordon coefficients over $M_A$, $M_1$ and over the total spins $S_{A1}$, $S_{B2}$ and $S$, A/B denote the spins on sub-lattices A and B and the spins on sites $1$ and $2$ make up the singlet. With A1(B2) we denote the set of spins on sublattice A(B) combined with spin 1(2). The sums can be evaluated and we obtain
\begin{eqnarray}
&&|\psi_{t<t_0}\rangle = \sum_{S=0}^{N-1} u^n_S 
{\scriptstyle \left( \sqrt{\frac{ (N-2+2 S)(N+4+2 S)  }{ 2 (N+2)^2  }} \left|  \frac{N+2}{4},\frac{N+2}{4},S,0\right\rangle \right.  }
\nonumber \\
&&{\scriptstyle +    
 \sqrt{\frac{ 2 S (S+1)  }{ (N+2)^2  }} \left| \frac{N+2}{4},\frac{N-2}{4},S,0\right\rangle }  
 {\scriptstyle - 
\sqrt{\frac{ 2 S(S+1)  }{ (N+2)^2  }} \left|\frac{N-2}{4},\frac{N+2}{4},S,0\right\rangle } \nonumber \\
&&{\scriptstyle +  
\left. \sqrt{\frac{ (N-2 S)(N+2+2 S)  }{ 2(N+2)^2  }} \left|\frac{N-2}{4},\frac{N-2}{4},S,0\right\rangle  
\right).} 
\end{eqnarray}

Again the equations simplify drastically in the continuum limit of large $N$, where as before $0 \ll S \ll N$. In this case the wavefunction of the system at $t<t_0$, expressed in the eigenstates of the $N+2$ spin system at $t=t_0$ is
\begin{eqnarray}
|\psi_{t<t_0}\rangle = \left[|0,n\rangle+|2,n\rangle \right] / \sqrt{2}.
\end{eqnarray}
Here all states on the right hand side, i.e. all the thin spectrum states labeled by their quantum number $n$ with either zero or two magnons, refer to configurations of $N+2$ spins.

To account for a finite temperature of our many particle qubit, we combine
initial states with different $n$ into a thermal mixture before we let it
interact with the two spin singlet. We should stress that we only consider
temperatures that are much below the magnon energy: $k_B T \ll J$ so that
there is no thermal occupation of the magnon states \footnote{Note that in a
  $d$-dimensional system the number of spins is $N=L^d$, where $L$ is the
  linear extent of the system. So the energy scale for the lowest possible
  spin wave (magnon) excitation is $J/L$. However the energy scale of the thin
  spectrum is $J/N=J/L^d$, and is thus --in any dimension higher than one--
  much lower than the magnon energy scale.}. 
This implies that the order
parameter is not affected by the thermal fluctuations. So, all that we
introduce is an incoherent mixture of the low lying thin spectrum states,
which all support a finite sublattice magnetization. Still, the implicit
assumption is that the thin states are in thermal equilibrium --and it is an
important assumption as our final result relies on it. In principle it can of
course not be excluded that occupation distribution of the thin spectrum
states is far from thermal equilibrium. But as we have not a priori prepared
the thin states in some particular way, we assume them to be thermally
occupied. Then the density matrix at times $t<t_0$ is then
\begin{eqnarray}
\rho_{t<t_0} 
&=& \frac{1}{Z} \sum_{n} e^{-\beta E^{n}_{0}} |0,n\rangle \otimes |singlet\rangle   
\langle singlet | \otimes \langle 0,n| \nonumber \\
&=& \frac{1}{2 Z} \sum_{n} e^{-\beta E^{n}_{0}}\left( |0,n\rangle\langle0,n| + |0,n\rangle\langle2,n| \right. 
\nonumber \\
&+& \left. |2,n\rangle\langle 0,n| + |2,n\rangle\langle2,n| \right) 
\end{eqnarray}
where, by definition, the partition function is $Z = \sum_{n} e^{-\beta E_0^n}$ and $\beta^{-1}=k_BT$. 

\section{Time evolution and decoherence.} 
By coupling the symmetry broken Lieb-Mattis model to the two spin singlet, we
have created the initial state of our $N+2$ spin qubit. This initial state is
precisely equivalent to the initial state~\eqref{eq:init} of the general description,
and we can thus follow equations~\eqref{eq:one}, \eqref{eq:two} and \eqref{eq:three} directly. That way we compute the
{\it exact} time evolution of the initial state density matrix, trace away the
thin spectrum states which have vanishing thermodynamic weight, and finally
define the off-diagonal element of the reduced density matrix as

\begin{eqnarray}
\rho^{OD}_{t>t_0} \equiv \frac {1}{Z} \sum_{n} e^{-\beta E^{n}_{0}}
e^{-i(E^{n}_{0}-E^{n}_{2}) (t-t_0)/\hbar}. 
\end{eqnarray}
We can then substitute the exact expressions for $E^m_n$ in this matrix element, and perform the summation. We find
\begin{eqnarray}
\rho^{OD}_{t>t_0}=\frac{1-e^{-x}}{1-e^{-N x}} \frac{1-e^{-N (x + i \tau)}}{1-e^{-x-i \tau}}
\label{eq:rhoOD}
\end{eqnarray}
where $x=\frac{E_{thin}}{k_B T} ~{\rm and}~ \tau=\frac{2}{\hbar}  \Delta E_{thin}
(t-t_0)$, with $E_{thin}=\sqrt{J B} ~{\rm and}~ \Delta E_{thin} =
E_{thin}/N$. We again define the coherence time $t_{spon}$ as the half-time of
$\left| \rho^{OD}_{t>t_0} \right|$. From the equation above one then finds
\begin{eqnarray}
t_{spon} \simeq \frac{2 \pi N \hbar}{k_B T},
\end{eqnarray}
our main result.

Notice that just as in the case of using a quantum crystal with an
interstitial excitation, the coherence time $t_{spon}$ in the end does not
depend on any details of the underlying model. The fact that $\Delta E_{thin}$
is proportional to $E_{thin}$ itself removes all dependence of $t_{spon}$ on
the model parameters $J$ and $B$. 

\subsection{Physical interpretation}
It is remarkable that the coherence time is such a universal time-scale, independent the detailed form of the thin spectrum --which, after all, is determined by the parameters $J$ and $B$ in the Lieb-Mattis Hamiltonian. Mathematically this is due to the fact that both $x$ and $\tau$ are proportional to $E_{thin}$. 
Physically one can think of this universal time-scale as arising from two separate ingredients. First, the energy of a thin spectrum state $|n\rangle$ changes when magnons appear, as we pointed out above.  The change is of the order of $n E_{thin}/N$, where $E_{thin}$ is the characteristic level spacing of the thin spectrum that we happen to be considering. The fact that each thin state shifts its energy somewhat at $t>t_0$ leads to a phase shift of each thin state and in general these phases interfere destructively, leading to dephasing and decoherence. The larger $n E_{thin}/N$, the faster this dynamics.

But from the argument above it is clear that in order for this dephasing to occur, it is necessary for a finite number of thin states to participate in the dynamics of decoherence. Since temperature is finite (but always small compared to the magnon energy) a finite part of the thin spectrum is available for the dynamics. Thin spectrum states with an excitation energy higher than $k_B T$ are suppressed exponentially due to their Boltzmann weights. Therefore the maximum number of thin states that do contribute is roughly determined by the condition that $n^{max} \sim k_B T/E_{thin}$. Putting the ingredients together, we find that the highest energy scale that is available to the system to decohere is approximately $ n^{max} E_{thin}/N =  \frac{k_B T}{E_{thin}} \frac{E_{thin}}{N}$. All together, the thin spectrum drops out of the equations. The time scale at which the dynamics take place is determined by the inverse of this energy scale, converted into time. From this argument we immediately find again the coherence time $t_{spon} \sim \frac{2 \pi \hbar N}{k_B T}$.

This physical picture also suggests an alternative way of introducing
decoherence into the many particle qubit. Instead of raising the temperature
and making an incoherent superposition of more and more thin spectrum states,
we could start out at $t<t_0$ with the Lieb-Mattis antiferromagnet in its
(zero temperature) symmetric ground state, and then instantaneously turn on
the symmetry breaking field $B$ at $t=t_0$. At $t>t_0$ the eigenstates are the
N\'eel-like thin spectrum states $|n\rangle$. No magnons are created by
switching on the symmetry breaking field. As we can expand the the states
$|n\rangle$ in the basis of total spin states as $|n\rangle = \sum_S u_S^n
|S\rangle$ we can, by the inverse transformation, expand the total spin
singlet state in the basis of the N\'eel-like thin spectrum states as
$|S=0\rangle = \sum_n u^0_n | n \rangle $. We can now use this singlet state
as the initial state for our qubit. This singlet state is a superposition of
all of the different N\'eel-like states, which are separated by energies
$E_{thin}$. This procedure thus roughly corresponds to creating a 'maximal
temperature' $k_B T \sim N E_{thin}$. When time evolves all of these states
pick up different phases, which leads to decoherence when we trace over them.
The coherence time due to this switching on of the symmetry breaking field is
therefore $t_{SB}=\frac{2 \pi \hbar}{\sqrt{JB}}$. 

\subsection{Symmetric case and short-ranged models}
This raises the question what would have happened if we had {\it not} broken
the symmetry in the Lieb-Mattis magnet (by introducing a finite symmetry
breaking field $B$) at all. In the symmetric case $S$, $S_A$ and $S_B$ are
good quantum numbers at all times. It is easy to see that in this situation
the thin spectrum, determined by the quantum number $S$, is independent of the
"magnon" states, which are determined by the quantum numbers $S_A$ and
$S_B$. Since in the symmetric Lieb-Mattis Hamiltonian the thin spectrum does
not communicate with the magnons and vice versa, we will find $\Delta
E_{thin}=0$, and accordingly no decoherence.

%The disappearance of decoherence in this setup is once again a demonstration of the singularity of the limit $B \to 0$: for all finite $B$ --where $B$ can be arbitrarily small-- we find the decoherence time $t_{spon}$, while the exactly symmetric system with $B=0$ will stay coherent forever.
%
The fact that $S$, $S_A$ and $S_B$ are all good quantum numbers, may be
regarded as a pathology of the Lieb-Mattis model. In fact, the model is
integrable just because there are so many conserved quantities. In a more
general, short ranged Heisenberg model the magnons will acquire a finite
lifetime and it is expected that they will in general influence the structure
of the thin spectrum, even if the symmetry breaking field is absent. In this
sense, the Lieb-Mattis model can really be seen as the best case scenario for
avoiding decoherence in non-Abelian, SU(2) symmetric models. Its infinitely
long ranged interactions introduce a large energy gap for all magnons, which
thus become extremely 'silent' excitations. On top of that the coupling to the
collective dynamics is so subtle that it can only be seen because of the
existence of a singular limit: Only because we need to always consider an
infinitesimal symmetry breaking field when looking at the thermodynamic limit
do we find decoherence at all.

%We checked that if we generalize the model somewhat by including a coupling between the magnons and the thin spectrum of the form $\frac{\eta}{N} m S^2$ (with $m$ the number of magnons present and the interaction strength $\eta$ is of order $J$), such a pathology does not occur. In this generalized model one finds $\tau=2 \eta t /(\hbar N^2)$. In that case one finds again that $t_{spon} \simeq \frac{2 \pi \hbar}{k_B T} \frac{J N^2}{\eta N} \sim \frac{2 \pi \hbar N }{k_B T}$.

\subsection{Recurrence}
Finally we notice that the off-diagonal elements of the density matrix, Eq.~(\ref{eq:rhoOD}), are periodic in time, and the initial density matrix recurs when $N \tau =2 \pi$ or, equivalently, $t_{rec}=\pi \hbar/E_{thin}$. Such a periodicity is required by the fact that the time evolution is unitary. As the recurrence time is inversely proportional to the level spacing of the thin spectrum, it depends on the microscopic parameters of the model. It becomes infinitely long if the symmetry breaking field vanishes. In the physical limit the recurrence time is always much longer than the decoherence time as $t_{rec}/t_{spon}=k_BT/E_{thin} \gg 1$.

\section{Conclusions}
A many-body qubit has an intrinsic limit to its maximum coherence time. In this paper we have presented in detail the considerations and calculations that lead to this conclusion. 

The limit to coherence is caused by the thin spectrum. In quantum systems a
continuous symmetry can spontaneously be broken in the thermodynamic limit due
to these states. The thin modes can be identified with the collective, zero
momentum, excitations of the orderparameter. In this paper we have outlined a general procedure for finding the thin spectrum states in a quantum systems.
The states within the thin spectrum are extremely low in energy and at the same time they are so few that their contribution to the thermodynamic partition function vanishes. 

If a symmetry breaking field is introduced, the resulting symmetry broken ground state is a superposition of (only) these thin spectrum states. The fact that the formation of the symmetry broken state occurs spontaneously in the thermodynamic limit can then easily be checked by considering the non-commuting limits of disappearing field and sending the number of involved particles to infinity. 

This has important consequences when a many-body quantum system is brought
into a superposition of two different internal states.
% a superposition of states with different number of elementary excitations
% --different number of magnons, for instance, in an antiferromagnet. In this
% case we have shown for the Lieb-Mattis antiferromagnet that 
We have shown that in that case the thin states will in general participate in
the time evolution of the full many-body system, even if their effect on any
thermodynamic quantity vanishes. This leads to dephasing and therefore
decoherence when the thin states are integrated out. We have found that the
time-scale corresponding to the dephasing process depends only on the energy
scale of the thin spectrum and the energy shifts induced in the thin spectrum
by the superposed initial states. Because the shifts in energy generally are
proportional to the level spacing itself, the decoherence time in the end
depends only on the temperature and size of the system, and not on the
underlying details of the model.

We have shown how  such superpositions can be defined and studied in a quantum
crystal and in the Lieb-Mattis antiferromagnet. The obvious question is to
what extent the Lieb-Mattis qubit is representative of a general many-body
qubit. In fact the Lieb-Mattis qubit is the best case scenario for the kind of
many-body qubits envisaged in main stream quantum information theory, as its
behavior is extremely close to semi-classical due to the presence of the
infinite range interactions. Qubits characterized by short range interactions
carry massless Goldstone modes and these will surely act as an additional heat
bath limiting the coherence time. It is of course not an accident that the
most 'silent' systems are qubits based on superconducting circuitry, which
have a massive excitation spectrum in common with the Lieb-Mattis system. We
have demonstrated here that even under these most favorable circumstances
quantum coherence eventually has to come to an end, because of the unavoidable
condition that even the most 'silent' qubits are subtly influenced by their
quantum origin. These effects become noticeable in the mesoscopic realm and
we present it as a challenge to the experimental community to measure the maximum
coherence time that they give rise to: $t_{spon} \sim \frac{2 \pi \hbar N}{k_B
  T}$.

\section{Appendix A: Thermodynamic weight of thin spectrum states}
It is easy to show that the contribution of the thin spectrum of the symmetric
$N$-spin Lieb-Mattis Hamiltonian to the free energy density is proportional to
$\frac{\ln N}{N}$ and thus vanishing in the thermodynamic limit. The energy of a state with total spin $S$ is $E_{thin}=J/N S(S+1)$ and its degeneracy is $2S+1$, so that the contribution of the thin states to the partition function is
\begin{eqnarray}
Z_{thin} &=& \sum_{S=0}^N (2S+1) \ e^{-\beta E_{thin}} \nonumber \\
&\approx& \int_0^N  (2S+1) \ e^{-\frac{\beta J}{N}S(S+1)} dS \approx \frac{N}{\beta J},
\end{eqnarray}
in the limit of large $N$. Therefore its total contribution to the free energy is $F_{thin}=-T\ln Z_{thin} \propto -\ln N$ and for large $N$ its contribution to the free energy per spin --and the free energy density-- is proportional to $\frac{\ln N}{N}$.

\section{Appendix B: Matrix elements}
By expressing the matrix elements of all components of the spin operators ${\bf S}_A$ and ${\bf S}_B$ in terms of Clebsch-Gordon coefficients, one can evaluate them by performing the appropriate summations. The resulting matrix elements are:

\begin{widetext}

\begin{eqnarray}
& \left< S_A' S_B' S' M' \left| S_A^{\pm} \right| S_A S_B S M \right> =  \nonumber\\
& \left[ \mp \delta_{S',S+1} \sqrt{\frac{\left(S'^2-\left(S_A-S_B\right)^2\right)\left(\left(S_A+S_B+1\right)^2-S'^2\right)\left(S'\pm M\right)\left(S'\pm M+1\right)}{4\left(4S'^2-1\right)S'^2}} \right. + \delta_{S',S} \frac{\left(\left(S_A-S_B\right)\left(S_A+S_B+1\right)+S\left(S+1\right)\right)\sqrt{\left(S \pm M+1\right)\left(S \mp M\right)}}{2S\left(S+1\right)} \nonumber \\
& \left. \pm \delta_{S',S-1} \sqrt{\frac{\left(S^2-\left(S_A-S_B\right)^2\right)\left(\left(S_A+S_B+1\right)^2-S^2\right)\left(S\mp M\right)\left(S\mp M-1\right)}{4\left(4S^2-1\right)S^2}} \right] \delta_{S_A',S_A}\delta_{S_B',S_B}\delta_{M',M \pm1}
\end{eqnarray}

\begin{eqnarray}
& \left< S_A' S_B' S' M' \left| S_A^{z} \right| S_A S_B S M \right> =  \nonumber  \\
& \left[ \delta_{S',S+1} \sqrt{\frac{\left(S'^2-\left(S_A-S_B\right)^2\right)\left(\left(S_A+S_B+1\right)^2-S'^2\right)\left(S'^2 - M^2\right)}{4\left(4S'^2-1\right)S'^2}} + \delta_{S',S} \frac{\left(\left(S_A-S_B\right)\left(S_A+S_B+1\right)+S\left(S+1\right)\right)M}{2S\left(S+1\right)} \right. \nonumber \\
& \left. + \delta_{S',S-1} \sqrt{\frac{\left(S^2-\left(S_A-S_B\right)^2\right)\left(\left(S_A+S_B+1\right)^2-S^2\right)\left(S^2- M^2\right)}{4\left(4S^2-1\right)S^2}} \right] \delta_{S_A',S_A}\delta_{S_B',S_B}\delta_{M',M}
\end{eqnarray}

\begin{eqnarray}
& \left< S_A' S_B' S' M' \left| S_B^{\pm} \right| S_A S_B S M \right> = 
- \left< S_A' S_B' S' M' \left| S_A^{\pm} \right| S_A S_B S M \right> +  \delta_{S',S} \sqrt{\left(S \pm M+1\right)\left(S \mp M\right)}   
\end{eqnarray}

\begin{eqnarray}
& \left< S_A' S_B' S' M' \left| S_B^{z} \right| S_A S_B S M \right> = - \left< S_A' S_B' S' M' \left| S_A^{z}\right| S_A S_B S M \right> +  M\delta_{S',S}.
\end{eqnarray}

\end{widetext}

\section{appendix C: The Bogoliubov Transformation}
We have used a Bogoliubov transformation to diagonalize bosonic bilinear Hamiltonians of the form
\begin{eqnarray}
H=\sum_k  A_k b^{\dagger}_k b^{\phantom\dagger}_k + \frac{B_k}{2} \left( b^{\phantom\dagger}_k b^{\phantom\dagger}_{-k} + b^{\dagger}_k b^{\dagger}_{-k} \right),
\end{eqnarray}
where $A_{-k}=A_k$ and $B_{-k}=B_k$.
The relevant transformed bosons $\beta^{\dagger}_k$ are defined through
\begin{eqnarray}
b^{\dagger}_k = \cosh(u_k) \beta^{\dagger}_{-k} - \sinh(u_k) \beta_k, \nonumber \\
b^{\phantom\dagger}_{-k} = \cosh(u_k) \beta_k - \sinh(u_k) \beta^{\dagger}_{-k}.
\end{eqnarray}
The parameters $u_k$ obey $u_k=u_{-k}$ and are chosen such that the Hamiltonian reduces to diagonal form. This implies that
\begin{eqnarray}
&&\cosh(2 u_k) = \frac{A_k}{\sqrt{A_k^2-B_k^2}}, \nonumber \\
&&\sinh(2 u_k) = \frac{B_k}{\sqrt{A_k^2-B_k^2}} ~~\text{and} \nonumber \\
&&H = \sum_k \sqrt{A_k^2-B_k^2} \left( \beta^{\dagger}_k \beta_k + \frac{1}{2} \right)- \frac{1}{2} A_k.
\end{eqnarray}
We can also use the exact same definition of $u_k$ to diagonalize the Hamiltonian
\begin{eqnarray}
H=\sum_k  A_k \left( a^{\dagger}_k a^{\phantom\dagger}_k +b^{\dagger}_k b^{\phantom\dagger}_k \right) + B_k \left( a^{\phantom\dagger}_k b^{\phantom\dagger}_{-k} + a^{\dagger}_k b^{\dagger}_{-k} \right) .
\end{eqnarray}
In that case the transformed bosons and Hamiltonian are given by
\begin{eqnarray}
a^{\dagger}_k &=& \cosh(u_k) \beta^{\dagger}_{-k} - \sinh(u_k) \alpha_k, \nonumber \\
b^{\dagger}_k &=& \cosh(u_k) \alpha^{\dagger}_{-k} - \sinh(u_k) \beta_k ~~\text{and} \nonumber \\
H &=& \sum_k  \sqrt{A_k^2-B_k^2} \left( \alpha^{\dagger}_k \alpha_k + \beta^{\dagger}_k \beta_k + 1 \right) - A_k.
\end{eqnarray}

\section{appendix D: The Order Parameter}
Consider the symmetry broken Lieb-Mattis Hamiltonian:
\begin{eqnarray}
H=\frac{2 J}{N}{\bf S}_A \cdot {\bf S}_B - B\left( S^z_A-S^z_B \right).
\end{eqnarray}
If the number of spins $N$ is large, then the eigenfunctions $\left|n\right>$ of this Hamiltonian are to a very good approximation given by the eigenfunctions of (half of) a harmonic oscillator:
\begin{eqnarray}
\left|n\right> &=& \sum_S u^n_S \left|S\right> \nonumber \\
&=& \sum_S \sqrt{\frac{\sqrt{\omega}}{\sqrt{\pi} 2^{n-1} n!} } e^{-\frac{1}{2}\omega S^2} H_n\left(\sqrt{\omega}S\right) \left|S\right>,
\end{eqnarray}
where $\left|S\right>$ are the total spin eigenstates, $H_n$ are the Hermite polynomials, $\omega$ equals $\frac{2}{N}\sqrt{\frac{J}{B}}$, and $n$ can only be an odd integer number. Using this exact expression to calculate the ground state expectation value of the order parameter, we find:
\begin{eqnarray}
&&\left< S^z_A-S^z_B \right> = \sum_{S,S'} u^1_S u^1_{S'} \left< S' \right|S^z_A-S^z_B\left| S \right> \nonumber \\
&&= 2\sum_S  u^1_S u^1_{S-1} \sqrt{ \frac{\left(\left(N/2+1\right)^2-S^2\right)S^2}{4 S^2 -1} }.
\end{eqnarray}
The shape of the function $u^1_S$ guarantees that $S \ll N$, so that for large $N$ the expectation value is approximately given by:
\begin{eqnarray}
&& \frac{N}{2} \int_1^{\infty} \sqrt{\frac{16}{\pi}}\omega^{3/2} S e^{-\frac{1}{2}\omega S^2} \left(S-1\right) e^{-\frac{1}{2}\omega \left(S-1\right)^2} dS \nonumber \\
&=& \frac{N}{2} \left[ e^{-\frac{1}{4}\omega} \left(1-\frac{1}{2}\omega\right)\left(1-\text{erf}\left(\sqrt{\frac{\omega}{4}}\right)\right)+\sqrt{\frac{\omega}{4}} e^{-\frac{1}{2}\omega} \right] \nonumber \\
&=& \frac{N}{2} e^{-\sqrt{\frac{J}{B}}\frac{1}{2 N}} + O(1),
\end{eqnarray}
which reduces to the classically expected order parameter in the thermodynamic limit. Note that in this expression it is immediately clear that the limit of vanishing symmetry breaking field does not commute with the limit of infinitely many spins: they form a singular limit.

\end{document}